\documentstyle[twocolumn,amssymb,aps,psfig,floats]{revtex}
\begin{document}
\draft

\twocolumn[\hsize\textwidth\columnwidth\hsize\csname @twocolumnfalse\endcsname

\title{Dynamical properties of the spin--Peierls compound 
$\alpha'$--NaV$_2$O$_5$}

\author{D.~Augier$^a$, D.~Poilblanc$^a$, S.~Haas$^b$, 
A.~Delia$^c$ and E.~Dagotto$^c$}

\address{
($a$) Laboratoire de Physique Quantique \& Unit\'e Mixte de 
Recherche CNRS 5626 \\
Universit\'e Paul Sabatier, 31062 Toulouse, France \\
($b$) Theoretische Physik, ETH H\"onggerberg, 8093 Z\"urich, Switzerland \\ 
($c$) Department of Physics and NHMFL, Florida State University,
Tallahassee, Florida 32306, USA
}

\date{April 97}
\maketitle

\begin{abstract}
\begin{center}
\parbox{14cm}{
Dynamical properties of the novel inorganic spin--Peierls compound 
$\alpha'$--NaV$_2$O$_5$ are investigated
using a one-dimensional dimerized Heisenberg model.
By exact diagonalizations of chains with up to $28$ sites, supplemented
by a finite-size scaling analysis, the dimerization
parameter $\delta$ is determined by requiring 
that the model reproduces the experimentally observed
spin gap $\Delta$.  The dynamical and static spin structure factors 
are calculated. As for CuGeO$_3$, the existence of a low energy
magnon branch  separated from the continuum is predicted.
The present calculations also suggest that 
a large magnetic Raman scattering intensity should appear above an 
energy threshold 
of $1.9 \ \Delta$.  
The predicted photoemission spectrum is qualitatively similar to results for
an undimerized chain due to the presence of sizable short--range
antiferromagnetic correlations.
}
\end{center}
\end{abstract}

\pacs{
\hspace{1.9cm}
}
\vskip2pc]

Recently, the quasi--one--dimensional (1D) compound
\hbox{$\alpha'$--NaV$_2$O$_5$} has received considerable attention
since it appears to be
the second inorganic material showing a spin--Peierls (SP) phase --- the
first one being CuGeO$_3$\cite{hase}.
Below a transition temperature $T_{SP} \approx 34K$, 
the compound undergoes a lattice 
distortion with the opening of a spin gap.
The structure of $\alpha'$--NaV$_2$O$_5$
is made of quasi--two dimensional layers of
VO$_5$ square pyramids separated by Na--ions~\cite{pouchard}. 
Two types of VO$_5$ chains alternate~: V$^{4+}$O$^{2-}_5$ 
and V$^{5+}$O$^{2-}_5$ (V$^{4+}$ carries
a spin $\frac{1}{2}$ while V$^{5+}$ does not). 
NaV$_2$O$_5$ is 
a good candidate for a 1D magnetic system since the magnetic
V$^{4+}$O$^{2-}_5$  
chains are isolated by non magnetic V$^{5+}$O$^{2-}_5$  
1D structures.

Originally, the presence of the SP phase transition was suggested by
experiments on polycrystalline samples\cite{isobe,ohama} which
showed a rapid reduction of the magnetic susceptibility below $T_{SP}
\simeq 34\ K$. From the dependence on the orientation of the magnetic
field, recent magnetic susceptibility measurements on single crystals 
unambiguously established the nature of the low--temperature phase
which is a  spin symmetric singlet ground state\cite{weiden}.  
The observation of structural distortions
by X--ray diffraction~\cite{fujii}, NMR~\cite{ohama} and 
Raman scattering\cite{weiden}
further suggested that an underlying spin--phonon coupling is 
responsible for the SP transition.
Note also that the critical temperature $T_{SP}$ is 
the highest of all known organic
or inorganic SP
compounds (e.g., $T_{SP}(\text{CuGeO$_3$}) \simeq 14\ K$).

The magnetic susceptibility $\chi(T)$ in the high--temperature
phase above $T_{SP}$
of this compound seems well--described by a 1D
antiferromagnetic (AF) Heisenberg model\cite{isobe,mila}.
Indeed, the crystallographic structure of this material suggests 
that the magnetic frustration is very small.
Fits of $\chi(T)$ below $T_{SP}$
provide estimates of both the nearest--neighbor spin exchange
$J$ and of the spin gap $\Delta$. 
Recent measurements on single crystals\cite{weiden} led to $J=441\ K$ 
and $\Delta \simeq 85\ K$ in good agreement 
with previous estimates of $J$~\cite{isobe,mila} and 
$\Delta$ \cite{ohama,fujii}. Then, in our analysis below
a ratio $\Delta/J=0.193$ is assumed. 

Following the approach used for the SP compound CuGeO$_3$, 
we will consider here a spin--1/2 AF Heisenberg model with
an explicit dimerization of the exchange coupling to account
for the lattice distortion,
\begin{equation}
H=J \sum_i (1+\delta (-1)^i)\: {{\vec{S}_i}\cdot{\vec{S}_{i+1}}} \ .
\label{hamiltonian}
\end{equation}
Note that in the case of CuGeO$_3$ an additional frustration 
was needed to describe the compound\cite{rieradobry,castilla}. 
The interchain couplings, although crucial to obtain a
finite ordering   temperature, are expected to
be  small and  will be neglected here.

\begin{figure}[hbt]
\begin{center}
\mbox{\psfig{figure=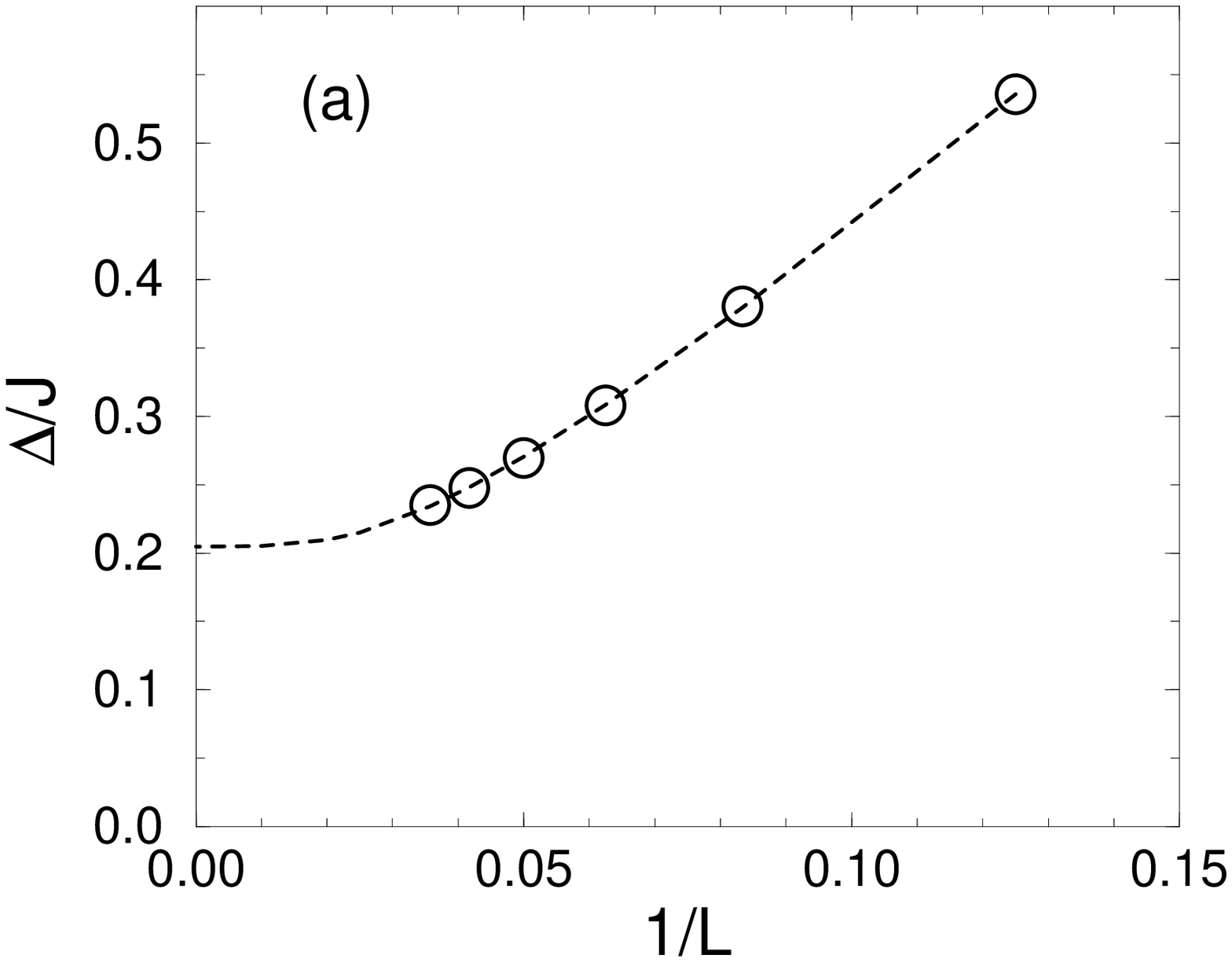,width=7cm,angle=0}}
\mbox{\psfig{figure=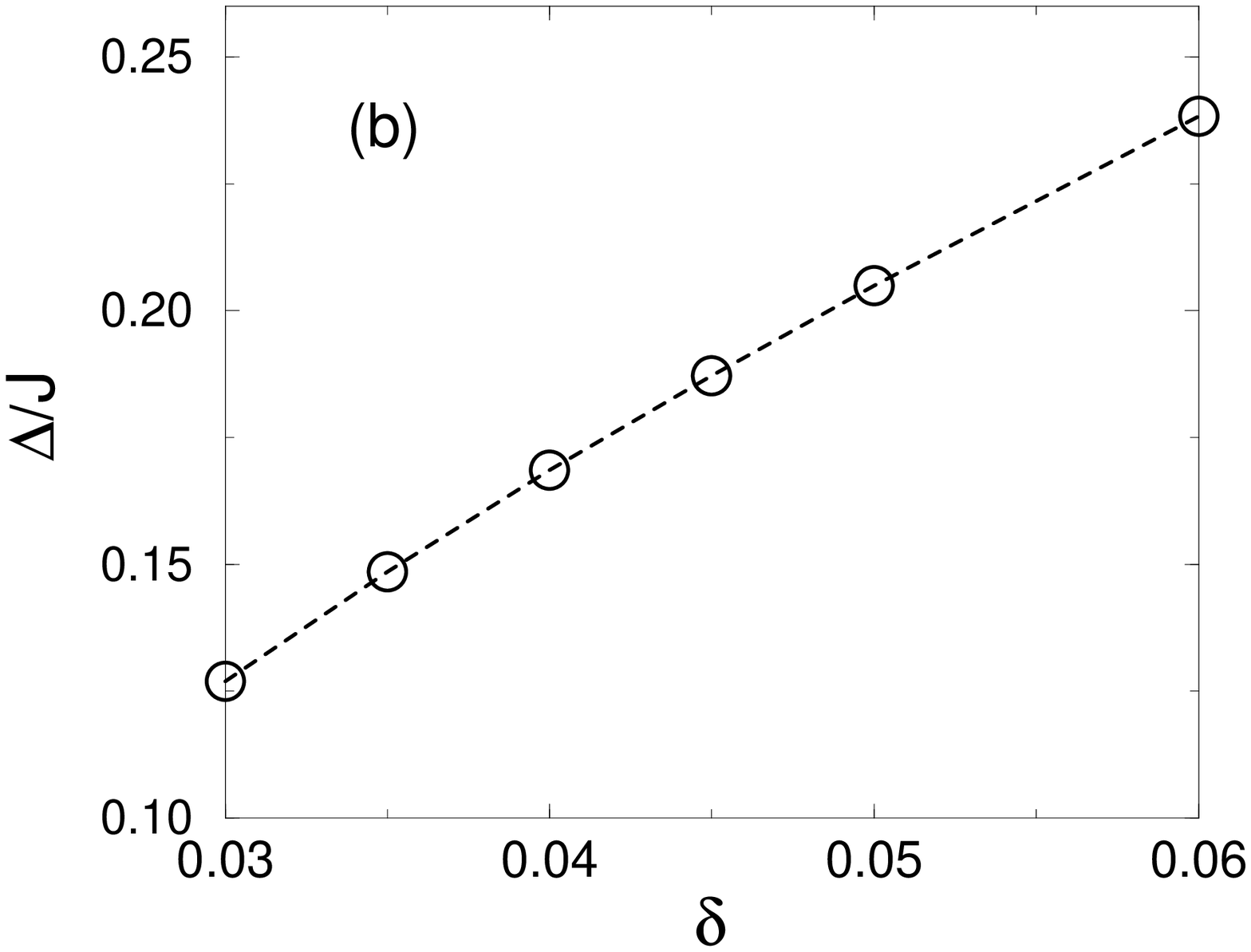,width=7cm,angle=0}}
\end{center}
\caption{(a) Spin gap $\Delta(L)$ in units of $J$
for $\delta=0.05$, as a function of 
$1/L$. The dashed line is the 
fitting curve described in the text. (b) Bulk extrapolated
value of $\Delta/J$ vs $\delta$.}
\label{scaling}
\end{figure}

The model Eq.(1) has a non--zero spin gap for all $\delta>0$,
and first we will determine the value of $\delta$ that
reproduces the experimentally measured  spin gap. The extrapolation
to an infinite chain $L\rightarrow \infty$ is performed accurately
using the scaling law
$\Delta(L)=\Delta+\frac{A}{L}\exp(-\frac{L}{L_0})$~\cite{bouzerar}.
The presence of a spin gap induces a length scale $L_0$ and 
finite size effects are negligible when $L \gg L_0$. 
As observed in Fig.~\ref{scaling}(a) for $\delta=0.05$,
this scaling behavior is indeed accurately satisfied. 
In addition, for such parameters we have found $L_0 \approx 18$ 
lattice spacing and, thus, extrapolations using data for systems with up to
$28$ sites are expected to have small error bars.
The behavior of the spin gap as a function of $\delta$ is shown 
in Fig.~\ref{scaling}(b). A comparison with the experimental 
value $\Delta/J=0.193$ gives an estimate  $\delta\simeq0.048$ for the actual 
NaV$_2$O$_5$ compound to be compared with $\delta \simeq 0.014$ 
obtained for CuGeO$_3$~\cite{rieradobry}. 
It is interesting to notice that the dimerization is larger for 
$\alpha'$--NaV$_2$O$_5$, although the ratios $\Delta/J$ are similar in 
both systems~\cite{note1}.  
The reason is that, in contrast to $\alpha'$--NaV$_2$O$_5$, 
a large frustration exists in CuGeO$_3$ from a sizable next--nearest 
neighbor coupling 
constant $J^\prime$. The frustration $J^\prime$ alone can 
produce a gap when the ratio $\alpha=J^\prime/J$ is larger than 
$\alpha_c\simeq 0.2411$.
For CuGeO$_3$ $\alpha\simeq0.36$ was proposed\cite{rieradobry}, while here
$\alpha=0$ is assumed.

Let us now proceed with the study of 
some dynamical properties of $\alpha'$--$NaV_2O_5$.
Typically, there are a number of experiments giving access to 
frequency dependent spectral functions of the form
\begin{equation}
I_A(\omega)\!=\!-\frac{1}{\pi}\lim_{\varepsilon\rightarrow0}\: \Im m \:
\langle\Psi_0|A\frac{1}{\omega+i\varepsilon-H+E_0}A^{\dagger}|\Psi_0\rangle,
\label{contfrac}
\end{equation}
where $\Psi_0$ is the ground state, $E_0$ its energy, and $A$ is some operator 
describing the physical process under consideration. Using 
exact diagonalization (ED) techniques,
$I_A(\omega)$ can be calculated with a continued fraction
expansion~\cite{ED}. An imaginary component $i\varepsilon$ is
added to $\omega$ in Eq.(\ref{contfrac}) providing a small width to the
$\delta$--functions.

\begin{figure}[hbt]
\begin{center}
\mbox{\psfig{figure=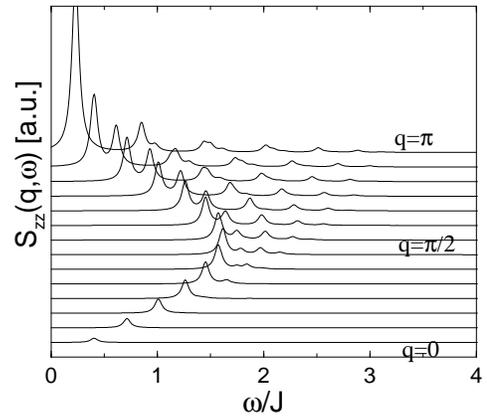,width=7.5cm,angle=0}}
\end{center}
\caption{Spectral function $S_{zz}(q,\omega)$ calculated 
at momenta $q=\frac{2\pi n}{L}$ ($L=28$ and $\delta=0.048$)
and using a small broadening $\varepsilon=0.04J$. From 
bottom to top, $n$ moves from $0$ to $14$.}
\label{szzqw}
\end{figure}

In particular, inelastic neutron scattering (INS) is an
accurate momentum dependent probe of the spin dynamics. 
Based on our model Eq.(1) and the parameter $\delta$ calculated here,
 we can predict the dynamical spin
structure factor $S_{zz}(q,\omega)$ measured by INS. 
$S_{zz}(q,\omega)$ is given by Eq.(\ref{contfrac}) with
$A=S_z(q)$ and $S_z(q)=1/\sqrt{L}\sum_j\exp(iqr_j)S_z(j)$.
The results on a 28 site chain 
are shown in Fig.~\ref{szzqw} for all momenta $q=n\frac{\pi}
{14}$, $n=0,\cdots,14$. We clearly observe a well--defined $q$--dependent 
low energy feature of bandwidth $\sim 1.6\ J$ having the largest
weight located around $q=\pi$. This is certainly reminiscent of the
Des Cloiseaux--Pearson~\cite{descloiseaux} (DP) excitation spectrum of the
Heisenberg chain. However, important differences arise from the presence of 
a spin gap: (i) there is no intensity for $\omega<\Delta$
at $q=0$ and $q=\pi$; (ii) the lowest singlet--triplet excitation branch
which has been interpreted as a spinon--spinon 
bound state~\cite{uhrig} is well separated from 
the continuum by a second gap.

\begin{figure}[hbt]
\begin{center}
\mbox{\psfig{figure=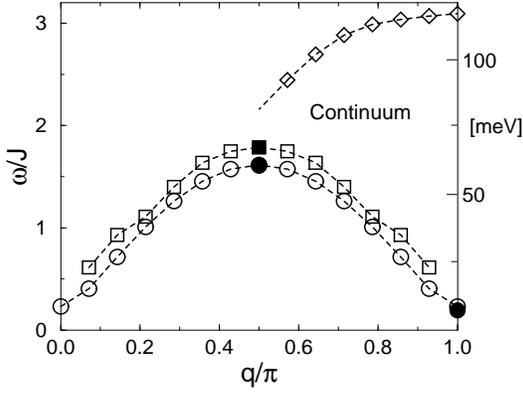,width=7cm,angle=0}}
\end{center}
\caption{Momentum dependence of the first
($\circ$),
second peak ($\square$)
and the upper limit for the continuum excitations ($\diamond$)  ($L=28$ and $\delta=0.048$).
The extrapolations to infinite size are also shown
for $q=\pi/2$ ($\blacksquare$, $\bullet$) and $q=\pi$ ($\bullet$). 
Units on the right are meV (assuming $J=440\ K$).}
\label{disp}
\end{figure}

The dispersion relation of the lowest energy magnon branch is presented 
in Fig.~\ref{disp}  with an infinite size 
extrapolation for momenta $q=\pi/2$ and $q=\pi$  (spin gap). The
finite size effects are quite small, especially for
$q=\pi/2$.  The second peak, as well as 
 the upper limit
of the continuum of excitations, are also shown. The dispersions of the
lowest excitations are  symmetric with respect to
$\pi/2$, reflecting the doubling of the unit cell by  dimerization. 
However, note that the spectral weight 
is {\it not} symmetric. As observed in Fig.~\ref{disp}, we have 
also explicitly 
checked at $q=\pi/2$ that the magnon excitation is
separated from the continuum by a gap.

\begin{figure}[hbt]
\begin{center}
\mbox{\psfig{figure=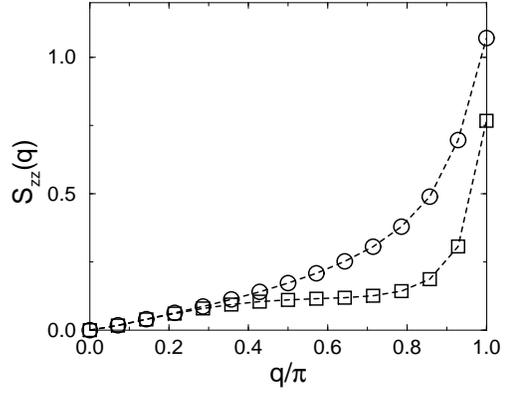,width=7cm,angle=0}}
\end{center}
\caption{Static structure factor $S_{zz}(q)$ ($\circ$) and weight of the
first excitation ($\square$) as a function of $q$ ($L=28$
and $\delta=0.048$).}
\label{stat}
\end{figure}

The static structure factor 
$S_{zz}(q)=\int d\omega\,S_{zz}(q,\omega)$
and the weight of the first peak are shown 
in Fig.~\ref{stat}. $S_{zz}(q)$ is sharply peaked
at $q=\pi$ due to strong 
short-range AF correlations. 
A sizable fraction of the weight is located above the magnon 
branch, specially at intermediate momenta such as  $q\simeq 5\pi/7$
where the continuum should be better observed experimentally.

It is interesting to make a quantitative comparison
with CuGeO$_3$. Assuming $J=440K$, Fig.~\ref{disp} 
shows that the maximum of the magnon branch occurs around 
$\omega_{max}\simeq 60$ meV while the spin gap 
is of order $\Delta\simeq 7.3$ meV.
Thus, the energy scales are approximately 4 times larger 
than for CuGeO$_3$~\cite{haas,poilblanc}
which might restrict the INS experiments to the bottom of the spectrum
around $q=\pi$~\cite{fujii}. 
Also, we observed that the ratio $\omega_{max}/J$ 
is close to the DP value of 1.57, while in CuGeO$_3$
it is approximately 1.2~\cite{poilblanc,regnault}.
This is due to the fact that the large frustration $J^\prime$ in CuGeO$_3$
affects the entire excitation spectrum, while in $\alpha'$--NaV$_2$O$_5$
only the low energy part of the spectrum is modified by the small 
dimerization. 
The ratio $\omega_{max}/J$ is then a key quantity to 
confirm experimentally  the absence of frustration in this system using INS.
In addition to the change in the value of $\omega_{max}/J$, frustration 
would also lead to a qualitatively different global structure 
of the spectrum. As example, the upper limit
of the continuum should be better defined for CuGeO$_3$ ($J^\prime \neq 0$) 
than 
for $\alpha'$--NaV$_2$O$_5$ ($J^\prime=0$). 

\begin{figure}[hbt]
\begin{center}
\mbox{\psfig{figure=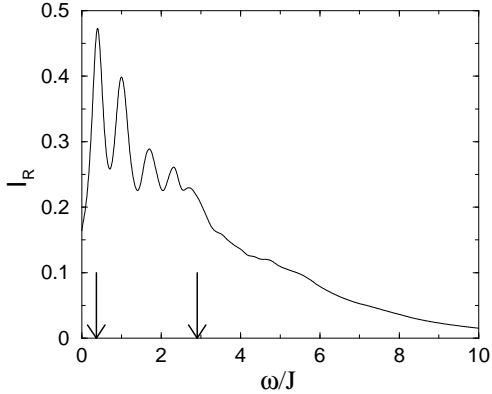,width=7cm,angle=0}}
\end{center}
\caption{Raman intensity for $L=28$ and $\delta=0.048$. 
A broadening $\varepsilon=0.2J$ was used.
The arrows indicate the extrapolated singlet--singlet gap $\Delta '$
at low energy and the first moment of the distribution $\langle \omega
\rangle$ at higher energy.}
\label{raman}
\end{figure}

Raman scattering is another powerful technique to 
probe the spin dynamics. 
The effective Hamiltonian for the photon--spin
interaction is given by~\cite{fleury}
\begin{equation}
H_{\mathrm{eff}}=g \sum_{<ij>}(\vec{e}_{in}.\vec{R}_{ij})
(\vec{e}_{out}.\vec{R}_{ij})\vec{S}_i.\vec{S}_j\ .
\end{equation}
$\vec{e}_{in}$ ($\vec{e}_{out}$) is the polarisation vector of the
incoming (outgoing) photons, the sum is over nearest--neighbor spins,
$\vec{R}_{ij}$ is the vector connecting them, and $g$ is a
coupling constant that depends on the
incoming photon frequency. 
$H_{\mathrm{eff}}$ is a spin--singlet, translationally invariant operator,
and it corresponds to physical processes involving the 
simultaneous excitations of two magnons with opposite momenta.
Since  the small interchain
coupling has been neglected, $\vec{R}_{ij}$ has to be collinear to
the chain. The largest Raman scattering intensity is thus 
expected for a 
polarisation of both photons along the chain 
direction. The Raman
operator can then be written~\cite{bouzerar,muth} (taking $g=1$)
as $H_R=\sum_i 
(-1)^i {{\vec{S}_i}\cdot{\vec{S}_{i+1}}}$
.
The Raman intensity $I_R(\omega)$ (Fig.~\ref{raman}) reveals a 
large scattering band centered at a mean energy (defined as the
first moment of the spectrum $\langle \omega\rangle=
\int d\omega \omega I_R(\omega)/\int d\omega I_R(\omega)$) of $\sim 2.9J$.
$I_R(\omega)$ is fairly smooth (the oscillations at low energy 
are finite size effects) and no Van Hove singularity 
is observed at the energy $2\ \omega_{max}$, associated with the 
top of the magnon branch. Due to the spin gap, an energy 
threshold $\Delta^\prime$ 
appears in the Raman scattering spectrum. The infinite size extrapolation 
of the corresponding singlet--singlet gap gives $\Delta'\simeq0.37J$
as indicated in Fig.~\ref{raman}.  
The ratio of the singlet--singlet gap over the singlet--triplet
gap is equal to $\Delta'/\Delta\simeq1.9$ close to the prediction of 
$\sqrt{3}$~\cite{uhrig}. Since we probe here 
double magnon excitations this suggests that 
magnons are almost non--interacting bosonic excitations (which would give
$\Delta'/\Delta=2$ exactly). 

\begin{figure}[hbt]
\begin{center}
\mbox{\psfig{figure=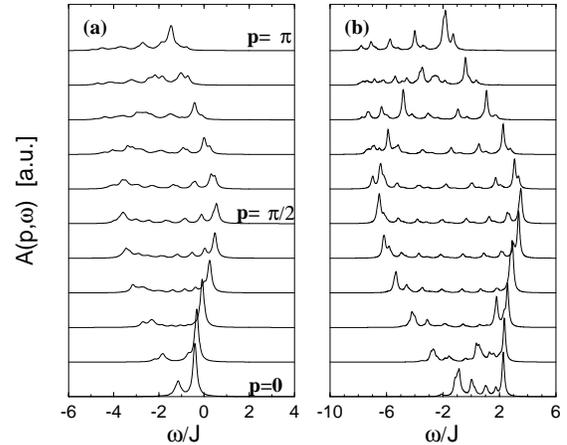,width=8cm,angle=-90}}
\end{center}
\caption{Hole spectral function $A(p,\omega)$ on a chain of 20 sites 
for $t/J=1$ (a) and $t/J=2.5$ (b) and broadening $\varepsilon=0.1J$. From 
bottom to top the momentum varies from 0 to $\pi$. 
}
\label{photo}
\end{figure}

We end our study of dynamical properties of $\alpha'$--NaV$_2$O$_5$
with an investigation of the
angle resolved photoemission spectrum (ARPES).
In this case the relevant operator $A$ in Eq.(\ref{contfrac})
is the destruction operator $c_{p\sigma}$ of an electron with momentum
$p$ and spin $\sigma$. Results will be presented for two
hole hopping amplitudes, i.e.
$t=J$ and $t=2.5\ J$, since it is {\it a priori} difficult 
to anticipate its actual value. 
The results shown in Fig. \ref{photo} are 
very similar to
the case of a single hole in a half--filled infinite--U Hubbard 
model~\cite{infinite_U}  or in the frustrated Heisenberg chain~\cite{haas}.
The overall scale of the spectrum is clearly given by $t$.
For $p<\pi/2$ a holon and spinon branches appear, while a 
``shadow band'' is observed for $p>\pi/2$ caused by short--range
 magnetic scattering 
at  $q=\pi$~\cite{haas}. 
Above the continuum, the high energy structure can be associated
with a reflection of the shadow band at the zone boundary\cite{infinite_U}. 
However, note that the presence of a spin gap introduces some subtle
differences with predictions for undimerized systems
that could be detected in ARPES experiments.
For instance, doping a spin gap insulator leads in general to a 
metallic state with only one zero--energy mode~\cite{ladder}
(corresponding to collective {\it charge} excitations only). 
This is reflected in 
Fig. \ref{photo} by the fact that the so--called spinon branch is in fact 
a broad structure instead of a branch--cut. In addition, the small peak 
at very low energy just above $k_F=\pi/2$  might be associated to a 
holon--spinon bound state. 
Similar results were obtained for CuGeO$_3$\cite{haas}. In spite of
these subtleties, it is clear that the spectral function of the
 dimerized model Eq.(1) has strong
similarities with undimerized systems. The main reason is that here an
explicit dimerization coexists with sizable short--range AF correlations,
a detail not sufficiently remarked in the literature on the subject.

In conclusion, using recent experimental data on the SP
$\alpha'$--NaV$_2$O$_5$ system,  the 
magnitude of the dimerization of the AF exchange coupling along the
chain has been determined.
In the framework of a 1D dimerized Heisenberg model, several
theoretical predictions for INS, Raman double
magnon scattering, and ARPES were here presented. 
Our calculations are expected to provide theoretical guidance for 
future experiments on SP systems.

We thank IDRIS (Orsay) 
for allocation of CPU time on the C94 and C98 CRAY supercomputers.
E. D. is supported by grant NSF-DMR-9520776.

\end{document}